\documentclass[final,5p,times,twocolumn, authoryear]{elsarticle}

\usepackage{amssymb}
\usepackage{booktabs} 
\usepackage{lipsum}
\usepackage[singlelinecheck=false]{caption}
\usepackage{makecell}
\usepackage{graphicx}
\usepackage{caption}
\usepackage{float}

\usepackage{hyperref}
\hypersetup{colorlinks,allcolors=black}

\makeatletter
\def\ps@pprintTitle{%
 \let\@oddhead\@empty
 \let\@evenhead\@empty
 \def\@oddfoot{}%
 \let\@evenfoot\@oddfoot}
\makeatother

\begin{document}
\begin{frontmatter}

\title{The Moment of Capture: \\How the First Seconds of a Speaker's Nonverbal and Verbal \\Performance Shapes Audience Judgments}

\author[a]{Ralf Schmälzle\fnref{label1}}
\affiliation[a]{organization={Department of Communication, Michigan State University},
            addressline={404 Wilson Rd.}, 
            city={East Lansing},
            postcode={48824}, 
            state={MI},
            country={USA}}
\affiliation[b]{organization={Brian Lamb School of Communication, Purdue University},
            addressline={100 North University St. }, 
            city={West Lafayette},
            postcode={47907}, 
            state={IN},
            country={USA}}
\author[a]{Yuetong Du}
\author[b]{Sue Lim}
\author[a]{Gary Bente}
\fntext[label1]{Corresponding Author. Email: schmaelz@msu.edu}

\begin{abstract}
Why do some speakers capture a room almost instantly while others fail to connect? The real-time architecture of audience engagement remains largely a black box. Here, we used motion-captured animations to present the pure nonverbal performance of public speakers to audiences - either in silence (nonverbal-only) or paired with the verbal content (nonverbal-plus-verbal). Using continuous response measurement (CRM), we find that audience judgments solidify with remarkable speed: Moment-to-moment engagement ratings become highly predictive of subsequent evaluations within the initial 10 seconds of the performance. Most notably, this predictive relationship emerged faster and slightly stronger in the nonverbal-only condition, with predictive information being present already after less than 5 seconds. These findings elucidate the social impact a speaker's nonverbal performance has on audience impressions, even when dissociated from the verbal content of the speech. Our approach provides a high-resolution temporal map of social impression formation, pointing to an early “moment of capture” that appears to set the stage for the reception of the following message. On a broader scale, this research validates a powerful new method to isolate different communicative channels, to scientifically deconstruct rhetorical skill, and to study the pervasive impact of nonverbal behavior more broadly. It also enables us to translate the ancient art of rhetoric into a modern science of social impression formation, yielding an empirical basis that can inform human-centered feedback, develop AI-based augmentation tools, and guide the design of engaging, socially present avatars in an increasingly AI-mediated and virtual world.
\end{abstract}


\end{frontmatter}

Imagine sitting in a conference session where a speaker enters the stage and starts the presentation. Within a short period of time, a subtle shift occurs: You may find yourself leaning forward, attentively engaged, and intent on hearing the rest; but it could also be that after only a few seconds, you start glancing at your phone or wander off into daydreaming. Similar reactions might occur not only in your mind, but also in those of others around you, pointing to a collective audience effect as the speaker either captures or loses the room. This collective engagement is a fundamental phenomenon in communication, a hard-to-pin-down, but very influential force that determines message impact. While we intuitively recognize this phenomenon, its elicitors and underlying processes remain unclear. What exactly happens in these initial, fleeting moments? To what extent is this due to the speakers’ confident and engaging body language, their opening words, or both? This paper presents a novel approach to measure the real-time impact of a speaker’s nonverbal and verbal performance on an audience.

In the following sections, we begin by laying out the theoretical foundation that bridges three research traditions: First, classical theories of rhetoric and research on audience engagement. Second, we connect this to nonverbal communication research and the social psychological literature on thin-slice impression formation, which highlights the critical importance of initial moments that capture audience attention. And third, we discuss methodological innovations that allow us to move beyond intuition and start to decipher the process of audience engagement on a moment-to-moment basis while controlling potential confounds. The current study uses high-precision motion capture animations and continuous response measures to test whether and from when on nonverbal 'thin slices' of scientific presenters can predict audience response.

\section*{Background}

\subsection*{The Rhetorical Tradition in Communication Research
 }
The study of public speaking is often described as the cradle of the communication discipline \citep{Peters1999,Sproule2014}. Rooted in classical rhetoric, analyses of public speaking have predominantly focused on the architecture of the message itself based on properties of the text (with the appeals to logos/logic, pathos/emotion, and ethos/credibility; 
e.g., \citealp{Aristotle2013,Cicero1942DeOratore,Cicero1949DeInventione}). This textual-interpretive tradition lends itself particularly well to understanding argumentation and composition in rhetoric (e.g., \citealp{Monroe1935,Wichelns1925}). However, with the rise of various new media (from radio and TV broadcasts, political propaganda and TV debates to entertaining Ted-Talks and popular video platforms like YouTube and TikTok), there has been what could be described as a “performative turn”, meaning that the speaker’s performance (or their mediated depiction) during delivery is seen as an important force determining psychological impact (e.g., \citealp{Wilke1934,frey1984nonverbale,MastersFreyBente1991,Druckman2003}). 

Speaking broadly, the speaker is not simply a vehicle to transmit the message content, but actively produces and manages the message by means of strategic choices as well as involuntary behaviors \citep{Goffman1956,Mehrabian1972,Haiman1949}. This has even been noted in classical rhetoric, where the concept of “\textit{actio}” (delivery) also features prominently. When looking at the components of messages that go beyond the purely textual information, perhaps the most obvious aspects are the speakers’ nonverbal behaviors (and paraverbal ones, see \citealp{Argyle1972,Hall1959,BurgoonDunbarSegrin2002}): A speaker’s posture, how they stand (static) and how they move (dynamic), their gestures, their management of space and interpersonal distance (e.g., distance from the audience, hiding behind the lectern, avoiding eye contact, or moving across the stage), and many other cues can strongly influence the impression on the audience – conveying feelings of warmth, competence, confidence, and so forth (e.g., \citealp{Mehrabian1969Posture,FiskeCuddyGlick2007}). In short, a speaker’s nonverbal behavior is a powerful tool in building rapport with an audience (e.g., \citealp{TickleDegnenRosenthal1990}) and communicating with “\textit{ethos}” or “\textit{charisma}” (e.g., \citealp{AndersenClevenger1963,Weber1922,Rosenthal1966}) during the live performance.

In this more performance-centric view of public speaking, the nonverbal channel is not just an ornament or peripheral cue: Rather, it is a rich channel that is readily available to the audience – perhaps more readily than the verbal channel itself. Moreover, the “silent code” of nonverbal behavior \citep{BurgoonManusovGuerrero2021,Mehrabian1972} is very immediate as it addresses audience members in an instinct-like way, and in a language that is largely shared across all humans; although there can be cultural differences in nonverbal communication, basic aspects related to sympathy, trust, and competence seem to be as universal as emotional expressions \citep{BenteEtAl2010Others,DarwinDarwin1872,FiskeCuddyGlick2007}. 

In short, by simply looking at a speaker’s nonverbal behavior – even if we don’t understand the content of the speech – audience members can make inferences about key social impressions, such as: Is this speaker confident? Do they have a firm grasp of their materials? Should I keep listening? On that view, the speaker’s body language could be said to deliver a meta-message about the message’s importance and the speaker’s conviction and competence in conveying the verbal message. Furthermore, it is clear that speakers themselves strategically exploit the nonverbal channel in their efforts to affect the audience (e.g., \citealp{MehrabianWilliams1969}). Some talented speakers, particularly in politics, are remarkably skilled at this (e.g., \citealp{Atkinson1984}); at the same time, a significant number of speakers also suffer from communication apprehension (e.g., \citealp{AllenBourhis1996}). 
\subsection*{Nonverbal Research on Impression Formation and Thin-Slices of Social Behavior
 }
A significant body of work in social psychology and nonverbal communication demonstrates that humans form rapid, durable, and often surprisingly accurate social impressions based on minimal information. In particular, an approach known as “thin-slicing” provides evidence for the speed and impact of first impressions \citep{AmbadyRosenthal1992}. For example, the classic thin-slicing study by Ambady and Rosenthal (\citeyear{AmbadyRosenthal1992})  showed that observers' judgments of classroom teachers, which were collected after viewing brief video clips of the teacher’s nonverbal behavior (under 30 seconds), were correlated with long-term outcomes, such as post-semester teaching evaluations. This suggested that a speaker’s core communicative style was reliably expressed within the video and consensually decodable by observers even after only very brief fleeting moments. 

Subsequent work has expanded this thin-slicing paradigm from teaching evaluations to other contexts, such as decoding the nonverbal behavior of doctors or lawyers \citep{Gladwell2005,SlepianBogartAmbady2014}. In general, it has been found that people readily form social impressions based on minimal information, as little as a static picture or a snippet of spoken text. Furthermore, these snap impressions tend to be surprisingly long-lasting and consequential – often predicting judgements and choices made after much longer periods of observation. While the validity of these impressions is not always clear \citep{Jussim2017}, there is no doubt that humans are uniquely equipped to pick up on social cues and use them heuristically to make consequential decisions (e.g., “\textit{I would hire you in a minute}”, \citealp{NguyenGaticaPerez2015}). 

Compared to the hundreds of thin-slicing studies in various social contexts,applications to the public speaking context are relatively rare, although some related work does exist. For instance, Chollet and colleagues (\citeyear{CholletOchsPelachaud2014,CholletScherer2017}) pioneered the use of computational methods to automatically analyze the public speaking performance from short clips. Similarly, \citet{CullenHarte2017} applied a thin-slicing style paradigm to TED-talks. However, it should be noted that the computational methods used in these earlier studies were mostly focused on easy-to-quantify concepts (e.g., pitch, motion energy), and the studies were done before the transformer-based breakthroughs in machine learning and in other contexts than science communication. 

More recently, \citet{BiancardiEtAl2025} presented a study of the so-called 3-minute-thesis (3MT) competition, which parallels our study because they were also interested in science communication. This study specifically investigated the timing of audience judgments, focusing on the beginning, middle, and end of the talk. Critically, they find a strong primacy effect: ratings of the initial slice were highly predictive of the talk winning the official audience prize. In contrast, the only other study we are aware of, done by \citet{GheorghiuCallanSkylark2020}, found a null result: 30-second video clips were not predictive of final evaluations. However, this study was done in the context of TED talks, which tend to be very engaging and speakers spend a lot of time to practice; therefore, it seems plausible that a ceiling effect may have occurred. Also closely related is a recent study by \citet{SchmalzleEtAl2025Audience}, which applied thin-slicing to science communication presentations. However, their study was exclusively based on the presentations’ textual transcripts, which were submitted as slices to various LLMs. Their findings suggested a strong thin-slicing effect (about 10-15 seconds of speech were enough to predict overall quality evaluations), but they sidestepped nonverbal and paraverbal cues altogether. The present study is designed to fill this gap by isolating the pure effect of nonverbal performance dynamics in the critical opening moments of a speech. In sum, while previous work clearly connects the thin-slicing paradigm from social psychology and education to the domain of public speaking, limited work exists in this area – and particularly the challenges associated with capturing, coding, and evaluating nonverbal performances, have hindered progress.

\subsection*{Studying Nonverbal Speaker Behavior and Measuring Audience Reactions: Challenges and Solutions
 }
Although the phenomena described above are intuitive and familiar to most people, they have presented at least three persistent methodological challenges. First, nonverbal communication behaviors are highly context-dependent and subtle, and they are best expressed in natural situations. However, many public speaking studies have used staged laboratory situations where speakers were instructed to exhibit certain behaviors, or they were introduced in ways to manipulate their ethos (or trustworthiness, source credibility, and related factors). This strategy has obvious benefits but also raises questions regarding external/ecological validity and confounds. Second, for studies that use videotaped performances (either staged or captured in real-life settings, etc.; \citealp{MehrabianWilliams1969,BiancardiEtAl2025}), the appearance of the speaker (i.e., static nonverbal information, such as attractiveness, race, age, or status cues) might interfere with or even overshadow any impressions arising due to their actual performance. This is reminiscent to the work on self-fulfilling prophecies which showed that manipulated expectations can influence impressions of the same behavior and reactions \citep{Jussim1986} . Third, human communication behavior involves a complex stream of verbal and nonverbal (and paraverbal) information, each of which can be further broken down into more molecular components. But this again introduces potential confounds: For instance, could a positive rating of a speaker’s performance just be driven by the fact that they presented about a more interesting topic or because it had significant results (e.g., \citealp{KahnemanFrederick2004})?

To overcome these challenges, our study combines two key methodological innovations. First, we use motion-captured animations on standardized avatars to isolate the speaker's nonverbal performance (e.g., \citealp{BenteEtAl2001}). This approach, rooted in classic biological motion studies (e.g., \citealp{Thornton2006}), provides unprecedented experimental control \citep{Bente1989,BenteEtAl2010Others,BenteEtAl2023Colocation}. Compared to classical video recordings used in similar contexts (e.g., \citealp{MehrabianWilliams1969,DuncanFiske1977}), which are prone to being biased by stereotypes (e.g., gender, race, attractiveness), the motion capture data can be used to render neutral, standardized avatars that then constitute a “pure” nonverbal stimulus.

Second, to capture the real-time formation of audience impressions, we employ Continuous Response Measurement (CRM). Instead of relying on a single, retrospective evaluation, CRM asks participants to continuously report their moment-to-moment evaluation of a stimulus. In political communication, for instance, \citet{IyengarJackmanHahn2016} found that partisan polarization in response to campaign ads can occur in under 30 seconds. In studies using virtual agents, \citet{CafaroEtAl2012} found that impressions of extroversion and friendliness formed within the first 12.5 seconds of interaction. 

\subsection*{The Current Study }
The present study leverages this theoretical framework and the discussed methodologies to conduct a direct empirical test of the thin-slicing principles in the context of public speaking. Our goal is to pinpoint the temporal dynamics of impression formation while seeking to disentangle the relative contributions of a speaker’s nonverbal performance from that of the verbal message. Using a corpus of speaker animations from motion-captured science presentations, this study has online observers provide both moment-to-moment engagement ratings (time-resolved measure of engagement) and retrospective summary evaluations (outcome measures of engagement). Moreover, these data were collected under two conditions. In a nonverbal-only condition, observers viewed and evaluated silent animations captured while speakers gave their talks; In the nonverbal-plus-verbal condition, they viewed the same animations accompanied by the audio channel information, i.e., the speech.

This allows us to identify the critical window of impression formation. Given prior work on the pervasiveness of first impressions and the speed of very short first slices (e.g., \citealp{WillisTodorov2006,SchmalzleEtAl2025Audience}), we expected that audience evaluations are formed very quickly and that nonverbal cues are a primary driver of initial impressions. 

\begin{figure*}[hbt!]
    \centering
	\includegraphics[width=0.8\textwidth]{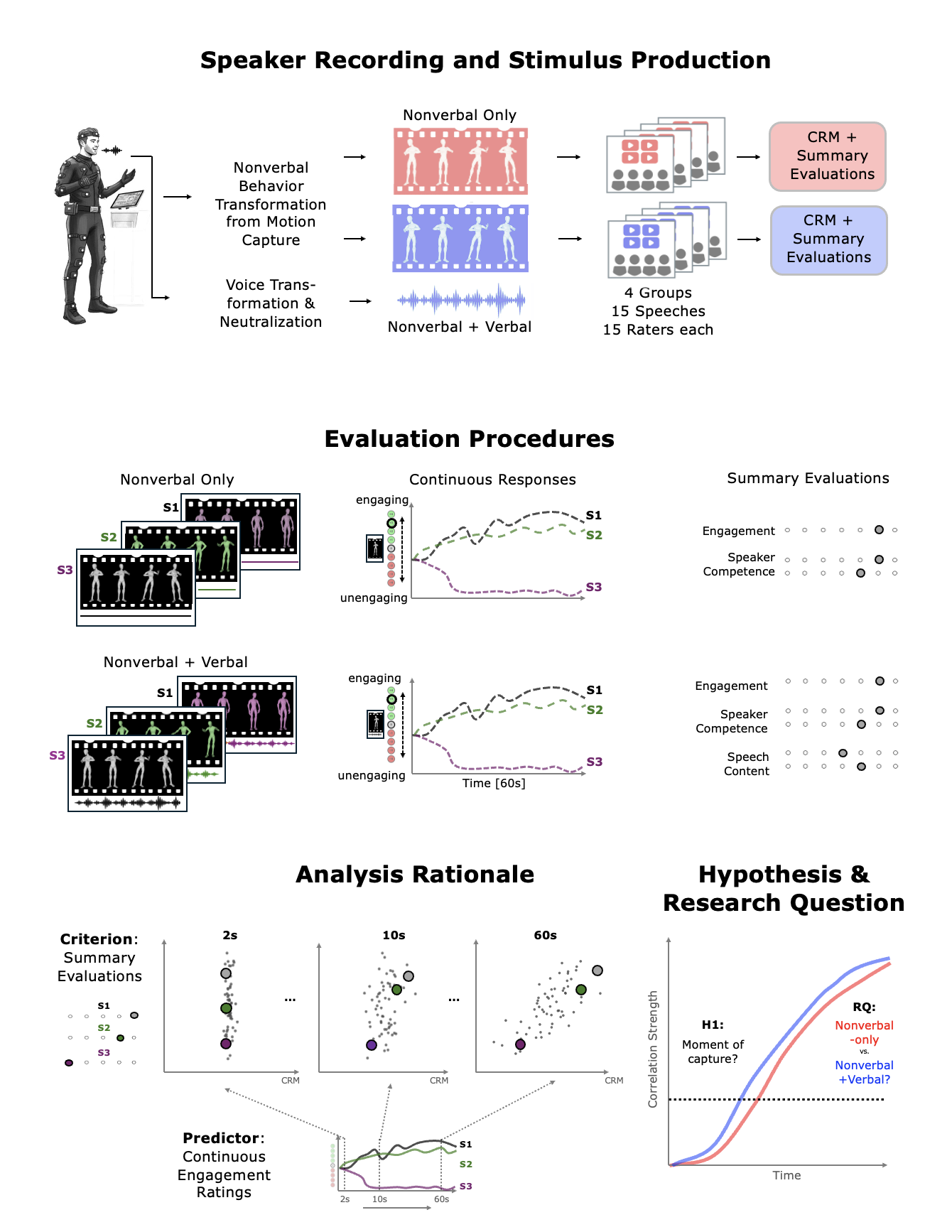}	
	\caption{\textbf{Figure 1. The Current Study: Stimuli, Conditions, and Analysis Overview.} Top panel: Illustration of stimulus creation based on real-life public speaking corpus, using motion-animation and voice-neutralization methods. A rating study was conducted to annotate these stimuli under either nonverbal-only or nonverbal-plus-verbal viewing conditions. Middle panel: Details on evaluation procedures: Nonverbal-only and nonverbal-plus-nonverbal performances were shown to test audiences, asking them to i) continuously evaluate engagingness and ii) provide retrospective summary evaluations (including engagingness as well as speaker and speech impressions). Bottom panel: Analysis rationale: The retrospective summary evaluations (collected after the performance) serve as the outcome criterion, and the continuously collected (CRM) ratings as the predictor. We then compute correlations between the moment-by-moment predictor values (CRM ratings at point1, point2, etc.) and the outcomes (final ratings). This enables us to map the temporal dynamics of impression formation and test at which point correlations are significant. Furthermore, we can disentangle the relative contributions of the nonverbal delivery from that of the verbal message.} 
	\label{fig_mom0}%
\end{figure*}

\textbf{H1}: Moment-to-moment engagement ratings (CRM) will be significantly predictive of retrospective evaluation. We expect that CRM ratings captured within the first 20 seconds of a speech will be strongly correlated with the final judgment (i.e., positively different from zero.)

\textbf{RQ1}: We expect that significant CRM-to-final-outcome correlations will emerge in both conditions – nonverbal-plus-verbal as well as nonverbal-only. Given the additional information conveyed via the verbal channel, it could be that the correlations in the nonverbal-plus-verbal condition rise faster than those in the nonverbal-only condition. However, it might also be that the nonverbal-only condition alone could contain enough information already. Thus, this research question explores how the part-to-whole thin-slice correlation time-courses evolve across the two conditions – nonverbal-only and nonverbal-plus-verbal, respectively.

\section*{Methods}
\subsection*{Study Overview}

To address these questions, we leveraged nonverbal and verbal data (high-fidelity body kinematics and speech recordings) from a large corpus of real-life public science presentations. Participants viewed these stimuli either in a nonverbal-only condition (seeing only the speaker animations) or a nonverbal-plus-verbal condition (seeing the speaker’s nonverbals and hearing their voice/speech) and provided moment-to-moment CRM ratings of engagement as well as retrospective summary ratings of engagement.  

\subsection*{Participants}
We recruited 120 human raters (\textit{mean\textsubscript{age}} = 40.3, \textit{sd} = 13.7, 58 self-identified males) to evaluate the speech performances. The study was conducted online via Qualtrics, using participants recruited via Prolific. The experimental protocol was approved by the local institutional review board; all raters provided informed consent and received monetary compensation for their participation in the study, which took about 35 minutes.

\subsection*{Stimuli and Procedures}
The experimental stimulus set consisted of 120 videos of 60 scientific presentations, with each presentation given in two conditions. In the nonverbal-only condition, participants just viewed the silent nonverbal avatar animations and were asked to provide CRM ratings of engagingness (see below); in the nonverbal-plus-verbal condition, participants viewed the same video, but were also able to hear the soundtrack of the speeches. The 60 videos in each condition were divided into four groups, for a total of eight groups (4*15 in the nonverbal-only, and 4*15 in the nonverbal-plus-verbal condition). Each of the eight groups of videos was viewed and rated by a group of distinct raters. This design was done to keep the time spent viewing and evaluating the speeches at a reasonable level and to avoid taxing participants’ limited attentional resources. 

\textit{\textbf{Stimulus Creation. }}The speeches, including all speakers’ body movements, were originally recorded as part of a larger study on public speaking \citep{LimSchmalzleBente2025}. Kinematic data were recorded using a high-precision Optitrack motion capture system (e.g., \citealp{BenteEtAl2001,BenteEtAl2025Clock}). Skeleton data was then extracted and rendered onto a set of standardized, emotionally and demographically neutral virtual avatars using the Vizard VR platform (see Figure 1). This process preserves the original speaker’s nonverbal behavior dynamics (e.g., posture shifts, swaying, illustrative and emphatic arm gestures) while systematically removing confounding visual identity cues such as attractiveness, gender, age, and ethnicity, which are known to strongly influence and bias social perception. Simultaneously, the original audio was transformed using ElevenLabs, a state-of-the-art AI-based voice cloning platform. We chose a gender-neutral voice as the target (a standard American English voice with a mid-range pitch; ElevenLabs Voice-ID = f5AWG6Xu8Fw3JCFUVWkS). This method preserved the exact verbal content and timing of the original speech while neutralizing paralinguistic cues (e.g., pitch, accent, vocal tone) that could otherwise signal stable speaker characteristics, such as age or gender.  Overall, this approach combines the ecological validity of naturalistic behavior research with experimental control, enabling us to zoom in on the constant, flowing stream of nonverbal movements and expressions while keeping body characteristics strictly controlled and adding/subtracting the voice track to offer/remove the speech content as context and to hold viewers’ attention.

\textbf{\textit{Rating Procedures of CRM and Outcome Ratings}}. While watching the presentations, participants were asked to use CRM to provide moment-by-moment evaluations of the speaker’s engagingness (see Figure 1). After viewing each presentation, participants provided a summary evaluation of engagingness. We chose engagingness as our primary outcome variable as it represents a holistic measure of the speaker's ability to establish and maintain an audience connection. Observers also provided additional ratings for overall rhetorical quality, speech interestingness (for those able to hear the verbal content), as well as two scales focusing on impressions of the speaker (5 items focusing on competence, charisma, nervousness) and impressions of the speech (interestingness of content, relevance, etc.). The CRM ratings were sampled at 2 Hz.

\subsection*{Statistical Strategies}
The two main sources of data consist of i) the CRM of engagingness and ii) the subsequently collected outcome evaluations for the same stimulus. The main question focuses on pinpointing the exact moment at which the CRM ratings become predictive of the outcome. Said differently, the strategy is to “slide” over the CRM ratings and correlate them to the outcome evaluations, keeping track of the resulting CRM-to-outcome correlations. This way, we can generate a time course of in-the-moment–vs–outcome-evaluations; the main question is when this time course crosses the significance threshold (see Figure 1, bottom panel). Second, we can also investigate how this time course evolves for the two viewing conditions (nonverbal-only vs. nonverbal-plus-verbal). Lastly, we use standard methods, such as averages, standard deviations, and correlations to examine, rank, and compare the different kinds of ratings to examine whether the different viewing conditions lead to similar results and how ratings for nonverbal-only viewing compare to ratings for nonverbal-plus-verbal viewing, as well as to ratings of the textual speech content alone (obtained from previous research).

\begin{figure*}[hbt!]
    \centering
	\includegraphics[width=0.8\textwidth]{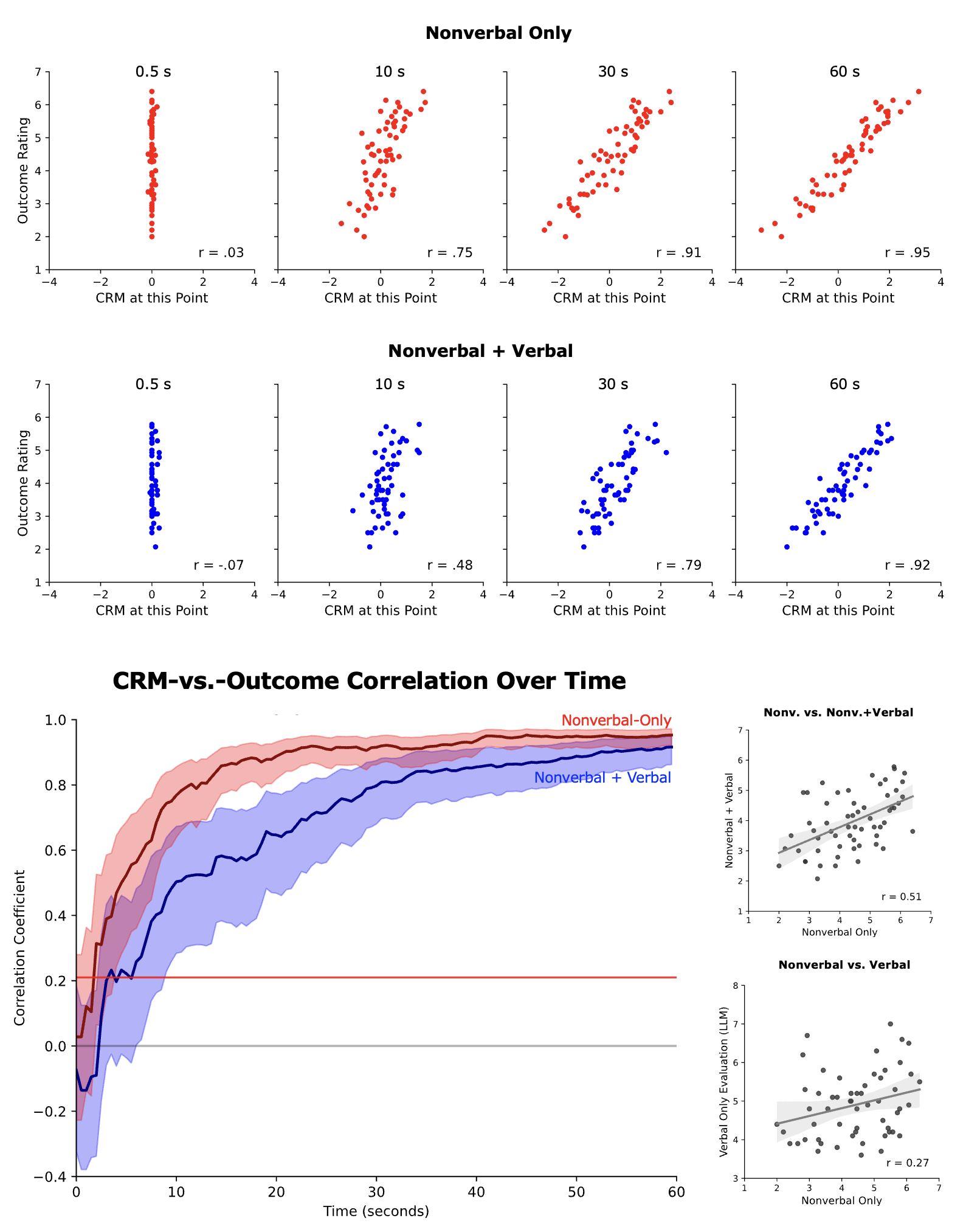}	
	\caption{\textbf{Figure 2. Correlations between CRM-Ratings and Subsequent Evaluations.} The top panels show the correlations between moment-by-moment CRM evaluations and the corresponding outcome evaluations for each speech – separated according to condition (red: nonverbal-only; blue: nonverbal-plus-verbal) and the CRM time points (0, 10, 30, and 60 seconds). The large panel shows the aggregated results (CRM-Engagement-Correlation curves) across all time points and for both conditions. As can be seen, the correlations between in-the-moment CRM ratings and final outcome ratings rise quickly, reaching the significance threshold (red line) already after a few seconds. Shaded areas represent the 95\%-confidence interval around the measured correlations for each time point. The small panels on the bottom show the positive relationship between nonverbal-only and nonverbal+verbal outcome engagement ratings (top) as well as the positive relationship between the nonverbal-only ratings and LLM-based evaluations of the textual transcripts of the same speeches (external to this study).} 
	\label{fig_mom0}%
\end{figure*}

\section*{Results}
\subsection*{Main Analysis: Thin-Slice (CRM) to Outcome Correlations}
To test H1, which stated that CRM ratings captured within the first 20 seconds of a speech should strongly predict final evaluations, we computed correlations between the moment-to-moment CRM ratings for every speech and the corresponding overall engagingness evaluation for these speeches, which were obtained after viewing. In other words, the outcome evaluation (engagingness rating obtained after viewing) was always the same, whereas the CRM values were based on a sliding analysis: We began with the CRM ratings at the speech onset, correlated them against the corresponding outcome evaluations and kept track of the results; then we shifted CRM ratings by a second and performed the same correlation against the after-speech ratings, again keeping track of the results, and so forth. The resulting moment-to-outcome correlation strengths were then plotted as a curve, and the same analysis was performed for both the nonverbal-only as well as the nonverbal-plus-verbal conditions.

As can be seen in Figure 2, the results strongly support H1: The correlations between the momentary CRM ratings and the outcome ratings of engagement rise very quickly from zero to asymptotic values of around 0.8-0.9. Moreover, this effect – the rapid rise and then convergence – is seen for both viewing conditions, the nonverbal-plus-verbal as well as the nonverbal-only condition. Also surprising was the speed at which the correlations rose. Using a significance threshold of \textit{r}(58) = 0.215 (\textit{p} < 0.05, one-tailed), it seems that useful information for predicting final ratings is already available in the first 3-5 seconds after speech onset; this is even faster than the 20 seconds we had hypothesized.
Regarding the research question, which strived to compare the evolution of the curves in Figure 2 across the nonverbal-only vs. the nonverbal-plus-verbal conditions, we observed the following: The nonverbal-only condition rose considerably faster and thus reached the critical threshold of \textit{r}(58) = 0.215 earlier. Moreover, the correlation between the CRM and outcome evaluations plateaued at a higher level (ca.\textit{ r} = 0.9), and they reached this plateau much earlier. In fact, after about 15 seconds into the stimulus, there is almost no additional information to be gained. By contrast, inspection of Figure 2 shows that for the nonverbal-plus-verbal condition, the correlations rise more slowly, only reaching the plateau of ca. \textit{r} = 0.8 after 30 seconds. Although the nonverbal-plus-nonverbal correlations are also quickly above zero, they linger between \textit{r} = 0.25 and \textit{r} = 0.8 for almost 40 seconds (i.e., from 5 seconds post-onset until about 45 seconds).

\subsection*{Additional Analyses}
In addition to these main analyses, we also explored the correlation between final outcome ratings on the same speeches in the two conditions. Between the nonverbal-only condition and the nonverbal-plus-verbal condition, the correlation lay at \textit{r} = 0.51, which is significantly positive (\textit{p} < 0.001; see Figure 2). Moreover, based on prior work examining LLM-based assessments of rhetorical quality (anonymized for review), we also had ratings (on a 1-10 scale) of rhetorical quality for the same speeches – but this time only for their textual transcripts. Thus, we are able to add another critical comparison: That between the engagingness of the nonverbal (silent language) performance and the corresponding evaluations for the same speeches’ text (verbal only content). Empirically, we find that a speaker’s “pure” nonverbal performance positively predicts the textual component with \textit{r} = 0.27, \textit{p} < 0.05 (see Figure 2).

\section*{Discussion}

This study examined whether and when early impressions of a public speaker’s nonverbal behavior can predict subsequent evaluations. To this end, we had participants give continuous evaluations of engagingness and correlated them with subsequent overall evaluations. Moreover, in one condition, participants received a full speech stimulus comprising both verbal content as well as how the content was delivered nonverbally, whereas another condition provided only the nonverbal delivery in isolation. We find that after less than five seconds into the speech, observers begin to become sensitive to social information that allows predicting subsequent evaluations of the speech. 

\subsection*{Discussion of Main Findings}
The most salient result of this work consists of the rapidly rising correlations (see Figure 2) between the moment-to-moment CRM ratings and the subsequent evaluations. This suggests two conclusions. First, relevant information, that is, information that allows valid forecasts of upcoming summary ratings, must be expressed by the speaker even within the very first seconds of a speech. Second, observers must be able to pick up on these cues. Of note, while this all relates to research done in a Brunswikian tradition \citep{Brunswik1955}, with notions of functional and ecological validity and cue-utilization, we did not link our findings to real-world outcomes (e.g., whether engaging speeches were also better remembered, etc.). Rather, the current results simply apply to the impressive speed with which people can extract social information from rather fleeting and very thin slices of social behavior, which relate to judgments made retrospectively and after much longer observation intervals.

The second key result worth discussing is that the rapid rise of part-to-whole correlations was also evident, and in fact even slightly faster and stronger, when they were only exposed to the speakers’ pure nonverbal behavior. Although this result aligns very well with the classical thin-slicing research done by Rosenthal and colleagues \citep{AmbadyRosenthal1992}, we found it still somewhat surprising because we had assumed that the nonverbal-plus-verbal condition would provide a richer medium with additional cues, which should have provided raters with better information to make faster ratings (e.g., \citealp{DaftLengel1986,Hall2001PONS}). Also, we had included the nonverbal-plus-verbal condition because we found that being able to “follow the speech” helped draw attention to the speaker compared to looking at the somewhat “meaningless movements” of the purely nonverbal avatar animations. However, contrary to our expectations, it turned out that observers were even slightly faster and obtained higher part-to-whole correlations when they saw only the nonverbal animations (particularly in the middle range, i.e., between 15-30 seconds, see Figure 1). Although speculative, it might be that the verbal content itself could be distracting, or that the intrinsic interestingness of the content might overshadow an otherwise dull performance. This would also fit with our observation that the spoken content of the first 30 seconds is usually highly formalized, typically comprising a greeting, a quick self-introduction, and perhaps a statement of the topic and an outline of the talk.

\subsection*{Theoretical Implications}

Our approach extends the thin-slicing methodology, and the main result, the rapidly rising correlation curves, bear resemblance to evidence accumulation dynamics (e.g., \citealp{GongHuskey2025}). Classical thin-slicing studies \citep{AmbadyRosenthal1992} showed that judgments from short clips predict outcomes. We here went a step further and provide – using well-controlled stimuli and in the context of science communication – a high-resolution map of how these impressions are formed in real-time as the speech unfolds. Inspired by the ‘histology’ metaphor \citep{AmbadyBernieriRicheson2000}, we used progressively thickening slices to document the microgenesis of thin-slicing with fine temporal resolution. 

One of the most interesting findings of the current research was that the nonverbal-only condition produced a faster and stronger prediction curve. This was not what we had expected, although there are several plausible reasons why this effect may have arisen: First, from an evolutionary and information-processing perspective, it could be that the nonverbal behavior may be processed more automatically and intuitively \citep{kahneman2003perspective}. This could be a powerful anchor, easy to decode, and highly consensual across observers. This would also be commensurate with the assumed strong effects of nonverbal communication (e.g., ethos/pathos, or in politics, etc.; e.g., Frey, 1982; \citealp{Argyle1972,Mehrabian1972}; but see \citealp{NagelMaurerReinemann2012} for a contradicting view in political rhetoric). 

However, our results, which show a rapid and strong predictive link between initial nonverbal cues and final judgments, appear to contrast with the null results reported by \citet{GheorghiuCallanSkylark2020} in a TED talk context. While this could be explained by the nature of their stimuli – TED talks, which were likely well-trained and pre-selected for higher quality – there exist also other studies with results that question a primary or dominant nonverbal influence: For instance, \citet{NagelMaurerReinemann2012}, using a CRM methodology similar to ours, analyzed a televised political debate and found that viewers' impressions were largely driven by the verbal discussion of political issues. In another elegant study, Jackob and colleagues (\citeyear{JackobRoessingPetersen2011}) manipulated a speaker's delivery of a fixed text. Their CRM data showed that while nonverbal delivery (vocal emphasis and gestures) could amplify the overall persuasiveness, the main audience reaction was tied to the verbal content.  
We propose this apparent divergence stems from methodological and contextual differences, which can easily harmonize the findings: First, the domain of politics is clearly different from our science communication context – not only in terms of topics, audience interest, and preexisting knowledge, but also in terms of audience involvement. Presumably, our audiences were less involved and thus nonverbal cues related to speaker dynamism had a much higher impact than in the context of politics, where audiences are more involved and hold pre-existing attitudes, and speakers are also more ideologically committed \citep{Dumitrescu2016}. Another possible explanation has to do with the relationship, integration, and mutual influence of verbal and nonverbal information, especially when considering that we only focused on initial impression formation. In the nonverbal-plus-verbal condition, the audience may indeed have a ‘richer’ stimulus, but also a potentially more confusing one: As argued above, especially in the first 30 seconds, the verbal information is often formulaic and low in novel information (greetings, repeating the title, showing the agenda). This low-value verbal information might actually distract from or dilute the high-value nonverbal cues, thus slowing down the judgment process. Thus, as also discussed by \citet{JackobRoessingPetersen2016}, it would be too simplistic to interpret our results as showing that the nonverbal channel dominates persuasive impact. After all, we did not even aim to study persuasion; our study’s contribution is not to refute or deny the importance of verbal content. However, we believe we have demonstrated a primary role of the nonverbal performance in the critical opening. It is this “moment of capture” that sets the stage for the reception of the following message. Thus, although we want to be careful to not overinterpret the findings, it seems plausible to argue that an audience’s assessment of “ethos”  is established through nonverbal “actio” (delivery), and this seems to be happening before “logos” (rational argument) unfolds. Viewed this way, the current empirical study is an empirical examination of this classical insight, providing a high-resolution, moment-to-moment map of how “actio” establishes a speaker's initial standing with an audience.

\subsection*{Practical Implications}
Moving from theoretical to applied implications, the current results have clear potential for public speaking training. In particular, they support the view that training should focus on mastering the first 15 seconds. Moreover, while much emphasis is placed on having attention-getting openers or using storytelling techniques to attract audience attention, it might be beneficial to explore the training of “nonverbal openers” and methods to increase speakers’ confidence and nonverbal immediacy. The rise of Virtual-Reality-based public speaking training, particularly combined with AI-powered feedback tools, seems very promising in this regard (see \citealp{CholletScherer2017,KroczekMuhlberger2023,VallsRatesNiebuhrPrieto2023,SaufnayEtAl2024}).

Related to this, as we enter a new era of AI- and metaverse-mediated communication (e.g., \citealp{ZionEtAl2025}), it will also be important that important nonverbal cues are mapped and transmitted appropriately. For instance, with the trend to online conferences and virtual social interaction (e.g., \citealp{AhnEtAl2024}), current avatar systems are not yet as expressive and flexible. However, recent advances in social AI and avatar technology, this is clearly on the horizon (e.g., \citealp{ZhangEtAl2025SocialAgent})

Finally, although our application context lies mainly in the domain of science communication and academic presentations (like the talks we see at scientific conferences), it is clear that the ability to engage audiences has consequential implications in various professional and societal contexts. For junior scientists, the most obvious implication is that they need to give a captivating job talk, but similar requirements also exist for business presentations or political and civil contexts. Thus, whether it is about landing a job offer,  being promoted, or winning elections, the ability to project charisma nonverbally is key to success. 

\subsection*{Strengths, Limitations, and Avenues for Future Research}
Strengths of this study include an unusually high level of experimental control, temporal precision, and a good tradeoff between internal and external validity. First, with motion-captured avatar behaviors, we were able to isolate the nonverbal dynamics from all other visual confounds (attractiveness, race, gender, age, attire). This allowed for one of the purest tests of the impact of movement possible. Second, using CRM offered a high temporal resolution, which allows to pinpoint the moment of capture in a way that retrospective ratings never could.  Lastly, although the avatar stimuli were precisely controlled and ‘neutralized’ in relevant ways, they still feature a high level of ecological validity as they stem from a corpus of real scientific presentations by practicing researchers and also preserve the naturalistic flow of bodily movement behavior, which is more natural and engaging to watch compared to, e.g., stick-figure or point-light walker animations.

Despite these strengths, a number of limitations should be kept in mind. Although the high level of control achieved allowed us to isolate the effects of body language in its purest form, this channel-isolation is also somewhat unnatural and removes other relevant channels, such as facial expressions or eye gaze, which could contribute independently or interact with body language to impact audience impressions. Second, although not necessarily a limitation, it should be kept in mind that the context of the study – scientific presentations – may not generalize to all public speaking situations (e.g., celebratory, political/persuasive, or classroom teaching situations beyond conference presentations, e.g., \citealp{benzion2025ai}). Finally, the engagingness construct is a holistic and overarching measure that clearly captures a basic evaluative dimension of public speaking. However, as with all verbal constructs and evaluative dimensions (e.g., \citealp{OsgoodSuciTannenbaum1957, FiskeCuddyGlick2007,SchmalzleEtAl2025Audience}), it would be interesting to compare the current results with other relevant aspects. For instance, while we asked audience subjects to evaluate engagingness continuously, the CRM literature in political communication usually focuses on agreement (e.g., \citealp{IyengarJackmanHahn2016}), and at least in a political communication context, it would seem relevant to also zoom in on this variable.

Turning from strengths and limitations to avenues for future research, we see the following areas as promising and viable. One unique strength of the avatar-animation technique is the possibility to reintroduce controlled information. For instance, it has been shown that it is possible to use the same nonverbal base-behavior to render it onto different avatars, as demonstrated e.g., by Bente and colleagues (\citeyear{BenteEtAl2010Others}) in an intercultural communication context. With this in mind, it would be possible to manipulate static appearance cues, such as female vs. male, professionally vs. casually dressed speakers, and measure how these impact judgments. Another promising avenue is physiological thin-slicing studies. While the CRM ratings provide a great way to study subjective perceptions over time, they come at a cost: Under normal circumstances, people would evaluate speakers implicitly, whereas the CRM method requires instruction \citep{Fahr2009reactivity}. Thus, we see great potential in translating the thin-slicing approach from the self-report/CRM domain to that of physiology. For instance, it could become feasible to examine audience-wide skin-conductance, heart-rate, or brain response in a similar fashion as shown here to understand how public speakers can induce and maintain collective audience responses. Finally, while the current study shows the potential of using thin-slices to predict long-term outcomes, we note that the long-term outcomes were still just final ratings. Going forward, it will be important to demonstrate whether this can also be linked to proximal outcomes. For instance, do these initial moments of capture predict better memory for the speeches, greater persuasion, or greater likelihood to be shared with others (e.g., \citealp{Falk2012getting,CoronelEtAl2021})?  

\subsection*{Summary and Conclusion}
This study addressed a core, longstanding question of communication: Why do some speakers capture an audience almost instantly, while others fail to connect? By isolating a speaker's pure nonverbal (body language) performance using motion-captured avatars and tracking audience reactions in real-time, we identified a critical “moment of capture”—a window of less than ten seconds where initial impressions solidify and strongly predict final judgments. Critically, this predictive relationship emerged even faster and more strongly when audiences were exposed to nonverbal behavior alone. In an increasingly mediated world, understanding how speakers capture (or lose) the audience is essential for training communicators, for deciphering the architecture of audience attention, and for understanding the nature of speaker-listener connection. 

\section*{Funding}
This study was supported by a grant from the National Science Foundation (Award No. 2302608: Neurocognitive and behavioral constituents of nonverbal speaker-listener attunement during science communication).

\section*{Data and Code availability}
Data and code are available via GitHub at \url {https://github.com/nomcomm/nonverbal_rating_ps}.

\bibliographystyle{elsarticle-harv}
\bibliography{main.bib}

\begin{figure*}[hbt!]
    \centering
	\includegraphics[width=0.6\textwidth]{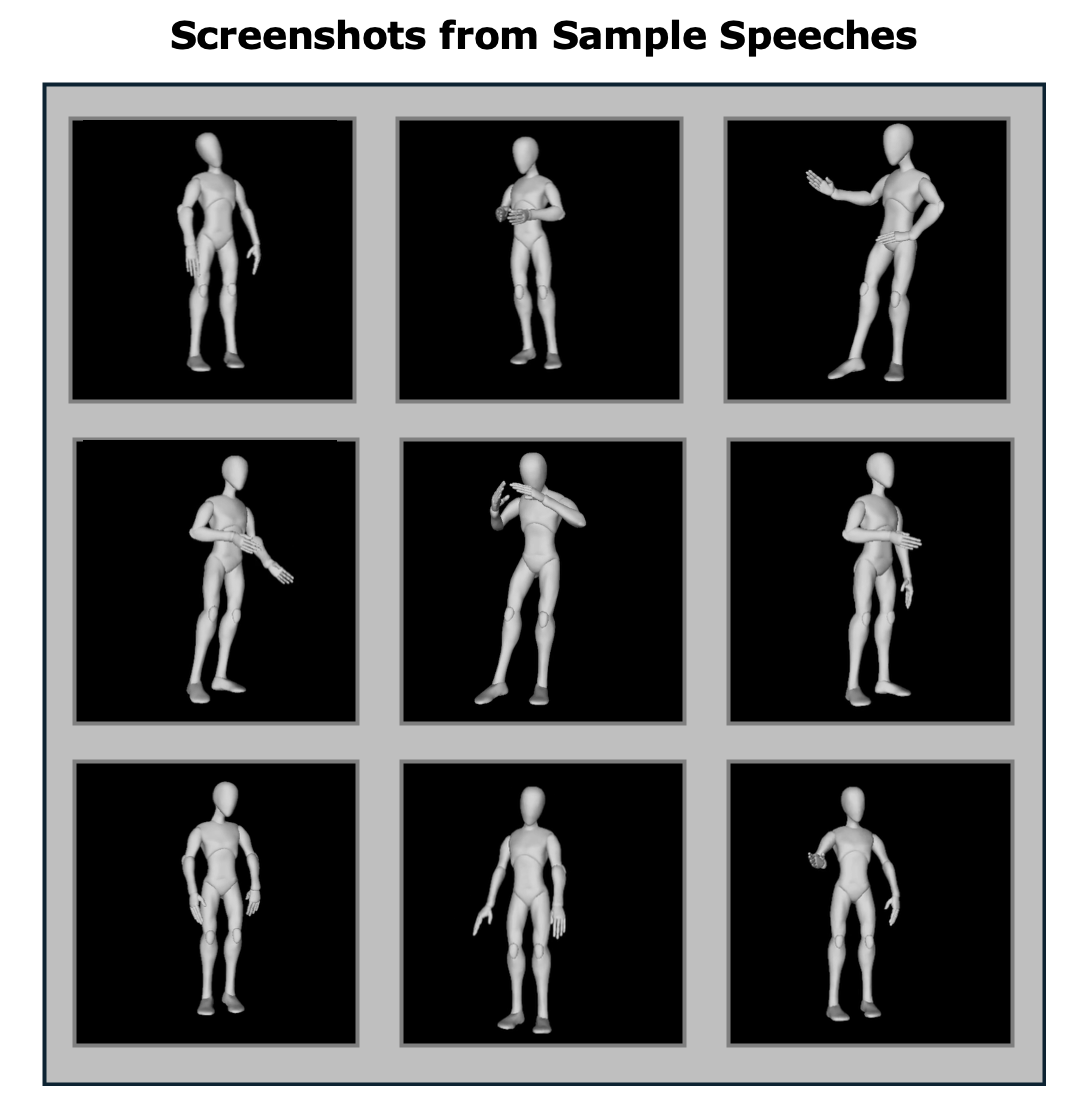}	
	\caption{\textbf{Supplementary Figure 1: Screenshots from nine different speeches, illustrating the nature of the nonverbal avatar animations.} Each video was 60 sec long and was generated based on the motion capture recordings from the real-life speakers presenting about their work, which was then re-rendered onto neutral avatar figures. This approach perfectly controls stereotypical information but preserves the continuous and natural flow of nonverbal behavior. Half of the observers watched and evaluated these silent animations; the other half watched the same animations accompanied by the corresponding verbal performance (also transformed in an analogous manner to neutralize vocal cues).} 
	\label{fig_mom0}%
\end{figure*}

\end{document}